\documentclass[]{raa}
\usepackage{graphicx}
\usepackage{times}             
\usepackage{amssymb}
\usepackage{amsmath}
\usepackage{mathrsfs}
\usepackage{float}
\usepackage{natbib}
\usepackage{booktabs}
\usepackage{hyperref}
\usepackage{xspace}

\begin{document}

   \title{Probing Dark Contents in Globular Clusters With Timing Effects of Pulsar Acceleration}

   \setcounter{page}{1}

   \author{Li-Chun Wang
      \inst{1,2}
  \and Yi Xie
      \inst{1} }


\institute{School of Science, Jimei University, Xiamen 361021, Fujian Province, China; {\it xieyi@jmu.edu.cn}\\
\and
Physics Experiment Center, Jimei University, Xiamen 361021, Fujian Province, China;}


\vs \no
   {\small Received 2020 Month Day; accepted 20XX Month Day}

\abstract{We investigate pulsar timing residuals due to the coupling effect of the pulsar transverse acceleration and the R$\rm{\ddot{o}}$mer delay. The effect is relatively small and usually negligible. Only for pulsars in globular clusters, it is possibly important. The maximum residual amplitude, which is from the pulsar near the surface of the core of the cluster, is about tens nanoseconds, and may hardly be identified for most of the globular clusters currently. However, an intermediate-mass black hole in the centre of a cluster can increase apparently the timing residual magnitudes. Particularly for the pulsars in the innermost core region, their residual magnitudes may be significant. The high-magnitude residuals, which above critical lines of each cluster, are strong evidences for the presence of a black hole or dark remnants of comparable total mass in the centre of the cluster. We also explored the timing effects of line-of-sight accelerations for the pulsars. The distribution of measured line-of-sight accelerations are simulated with a Monte Carlo method. A two-dimensional Kolmogorov-Smirnov tests are performed to reexamine the consistency of distributions of the simulated and reported data for various values of parameters of the clusters. It is shown that the structure parameters of Terzan 5 can be well constrained by comparing the distribution of measured line-of-sight accelerations with the distributions from Monte Carlo simulations. We provide that the cluster has an upper limit on the central black hole/dark remnant mass of $ \sim 6000 M_{\odot} $.
}

   \authorrunning{WANG \& XIE}            
   \titlerunning{Probing Dark Contents in GCs With Pulsars}  
   \maketitle

\section{Introduction}\label{sec:intro}

Intermediate-mass black holes (IMBHs) are black holes with mass between $10^2-10^5~M_{\odot}$. They are considered as the missing link between stellar black holes and supermassive black holes \citep{2013snpa.confE...2H}. Although it has long been suspected that IMBHs may form in centres of globular clusters (GCs) \citep{1975Natur.256...23B}, the existence of IMBHs in GCs remains in doubt.

Two traditional methods have been widely used to reveal the IMBHs in GCs. The first is studying the dynamics of the stars through optical observations, by which the past researches inferred upper limits on the mass of the IMBHs \citep{2006ApJS..166..249M,2010ApJ...710.1063V,2020MNRAS.499.4646A}, or made a few tentative detections \citep{2008ApJ...676.1008N,2013A&A...552A..49L}. The second is looking for signatures of X-ray and radio emission from the accreting IMBHs \citep{2004MNRAS.351.1049M}. However, some controversial limits of the masses were obtained with this approach \citep{2006ApJ...644L..45P,2006MNRAS.368L..43D,2012ApJ...755L...1M,2013ApJ...776..118S,2018ApJ...862...16T}. Recently, an IMBH in an extragalactic stellar cluster might have been found through the observation of a tidal disruption event \citep{2018MNRAS.474.3000L}.

Pulsars are very stable rotators, which emit radio pulses to the Earth with regular arrival times. Pulsar timing analysis is based on the measurement of the precise times of arrival (TOAs) at the telescope, which provides the spin period and its derivatives of the pulsar. However, many pulsars exhibit significant timing irregularities, i.e., unpredicted arrival times of the pulses, namely timing residuals. Pulsar timing residuals, which are widely exist, can be produced by many mechanisms \citep[e.g.]{2010MNRAS.402.1027H}. One of which specially concerned here is an observed delay, the R$\rm{\ddot{o}}$mer delay, due to the Earth being at a point of its orbit further from the pulsar as compared to the times when it is at the point nearer the pulsar. Then, a growing sinusoidal pattern of the residual will be observed if the pulsar has proper motion. This was first achieved for PSR B1133+16, using a four-year period of observations \citep{1974ApJ...189L.119M}. The coupling effect of the pulsar transverse acceleration $a_{\perp}$ and the R$\rm{\ddot{o}}$mer delay induces very similar timing residual patterns, but their oscillation envelopes consists of two parabolic curves, rather than straight lines \citep{2020RAA....20..191X}. However, the residual is usually negligible in timing analysis, since it is relatively small \citep{2006MNRAS.372.1549E}, and at present there is no relevant report for it .

GCs are extremely prolific MSP factories. Observational and theoretical evidences suggest that they are probably more than $\sim 1000$ MSPs in some massive GCs \citep{2013MNRAS.436.3720T}. It was anticipated that a pulsar in dense environment of the host GC would experience varying accelerations due to close encounters with nearby stars, and this effect could be used as a probe of the cluster dynamics \citep{1987MNRAS.225P..51B}. It was also noticed that the resultant perturbations in acceleration cannot be measured directly using pulsar timing analysis, since they would be absorbed into the pulsar spin parameters of the timing model \citep{1987MNRAS.225P..51B,1992RSPTA.341...39P}. However, due to the motion of a pulsar through the gravitational potential of GCs, the Doppler effect at the line-of-sight direction may overwhelm the intrinsic positive spin period derivative $\dot P$, the concept was further developed to study the intrinsic characteristics, structure and components of the GCs \citep{1992RSPTA.341...39P,1993ASPC...50..141P}. Very recently, new evidences were obtained for the identification of the IMBHs, by measuring its effects on accelerations, jerks or jounces of MSPs in GCs \citep{2017ApJ...845..148P,2017MNRAS.471..857F,2017MNRAS.468.2114P,2019MNRAS.487..769A, 2019ApJ...884L...9A}. Further, the pulsar acceleration measurements together with the dynamical N-body simulations of the cluster, provided some striking constraints on the mass of an IMBH in the centre of the GC \citep{2017Natur.542..203K,2017MNRAS.464.2174B,2019MNRAS.488.5340B}.

However, only the line-of-sight acceleration $a_{\rm l}$ of pulsars is concerned in the previous measurements for the IMBHs. The effect of the transverse acceleration $a_{\perp}$, which contains complementary information, is still received little attention. It was proposed that, the magnitudes of the timing residuals due to the coupling effect of $a_{\perp}$ and the R$\rm \ddot{o}$mer delay, may be important for pulsars deposited to the core region of a nearby GC \citep{2020RAA....20..191X}. Probably, if an IMBH presents in the centre of a GC, $a_{\perp}$ may be sensitive to the mass of the black hole, especially for the pulsars in inner region. Thus, this type of timing residuals (due to the coupling effect) may possibly be identified for pulsars in the centre region of the GC. In turn, the amplitudes of the residuals may also provide additional constraints on the mass of the central black hole.

In this work, we estimate the magnitudes of the residuals for pulsars near the core region of GCs in section 2, the residuals can be used to probe the mass of IMBHs in the centre of GCs, and the structure parameters of Terzan 5 can be constrained by comparing the distribution of measured $a_{\rm l}$ with the distributions from Monte Carlo simulations, as shown in section 3. Finally, the results are summarized and discussed in section 4.

\section{Methods}

The coupling effect is essentially attributed to the geometric propagation delay. Following the convention of the math in timing analysis \citep{2006MNRAS.372.1549E}, the displacement $\boldsymbol{k}$ of a pulsar may be broken into the first and second derivatives,
\begin{equation}\label{displacement}
\boldsymbol{k} = \boldsymbol{\mu} |\boldsymbol{R_{0}}|(t^{\rm psr}-t_{\rm pos}) +\frac{\boldsymbol{a}}{2}(t^{\rm psr}-t_{\rm pos})^2,
\end{equation}
where $\boldsymbol{\mu}$ is the velocity divided by the distance, $\boldsymbol{R_{0}}$ is the position vector of the pulsar, $\boldsymbol{a}$ is the acceleration vector, $t^{\rm psr}$ is the proper time measured at the pulsar since epoch $t_{\rm pos}$. Substituting the displacement $\boldsymbol{k}$ into the annual proper motion term, i.e. the second term in the first pair of parentheses of equation (5) in \cite{2006MNRAS.372.1549E}, we have
\begin{equation}\label{annual proper motion}
\frac{\boldsymbol{k}_{\perp}\cdot\boldsymbol{r}_{\perp}}{|\boldsymbol{R_{0}}|}=\boldsymbol{\mu}_{\perp}\cdot \boldsymbol{r}_{\perp} (t^{\rm psr}-t_{\rm pos})+ \frac{\boldsymbol{a}_{\perp}\cdot\boldsymbol{r}_{\perp}}{2 |\boldsymbol{R_{0}}|}(t^{\rm psr}-t_{\rm pos})^2,
\end{equation}
in which $\boldsymbol{r}$ is the barycentric position of the observatory, the radial and transverse components are denoted by subscripts, i.e. $i_{\parallel}=\boldsymbol{i}\cdot \boldsymbol{R_{0}}/|\boldsymbol{R_{0}}|$ and $\boldsymbol{i}_{\perp}=\boldsymbol{i}-i_{\parallel}\boldsymbol{R_{0}}/|\boldsymbol{R_{0}}|$, and $\boldsymbol{i}$ is an arbitrary vector. In equation (5) of \cite{2006MNRAS.372.1549E}, only the first three terms in the first pair of parentheses involve the displacement $\boldsymbol{k}$ of the pulsar. The first term corresponds to the Shklovskii effect, which mainly acts on the line-of-sight acceleration and has no other important effect on the timing residual, is a tiny contribution to the measured acceleration \citep{2017ApJ...845..148P}, and is also small compared to the second term, since $|\mathbf{k}|\ll |\mathbf{r}|$. The third term is ignored, since it is analogical to the second term, and $|\boldsymbol{b}|\ll |\boldsymbol{r}|$ for almost all the binary pulsars in GCs, where $\boldsymbol{b}$ is the position of the pulsar with respect to the binary barycentre. The absolute values of the second, third and forth terms in the second pair of parentheses are much smaller than unit. Thus only the second term in the first pair of parentheses is considered in this work.

\begin{figure*}[!htbp]
\centering
\includegraphics[width=0.95\textwidth]{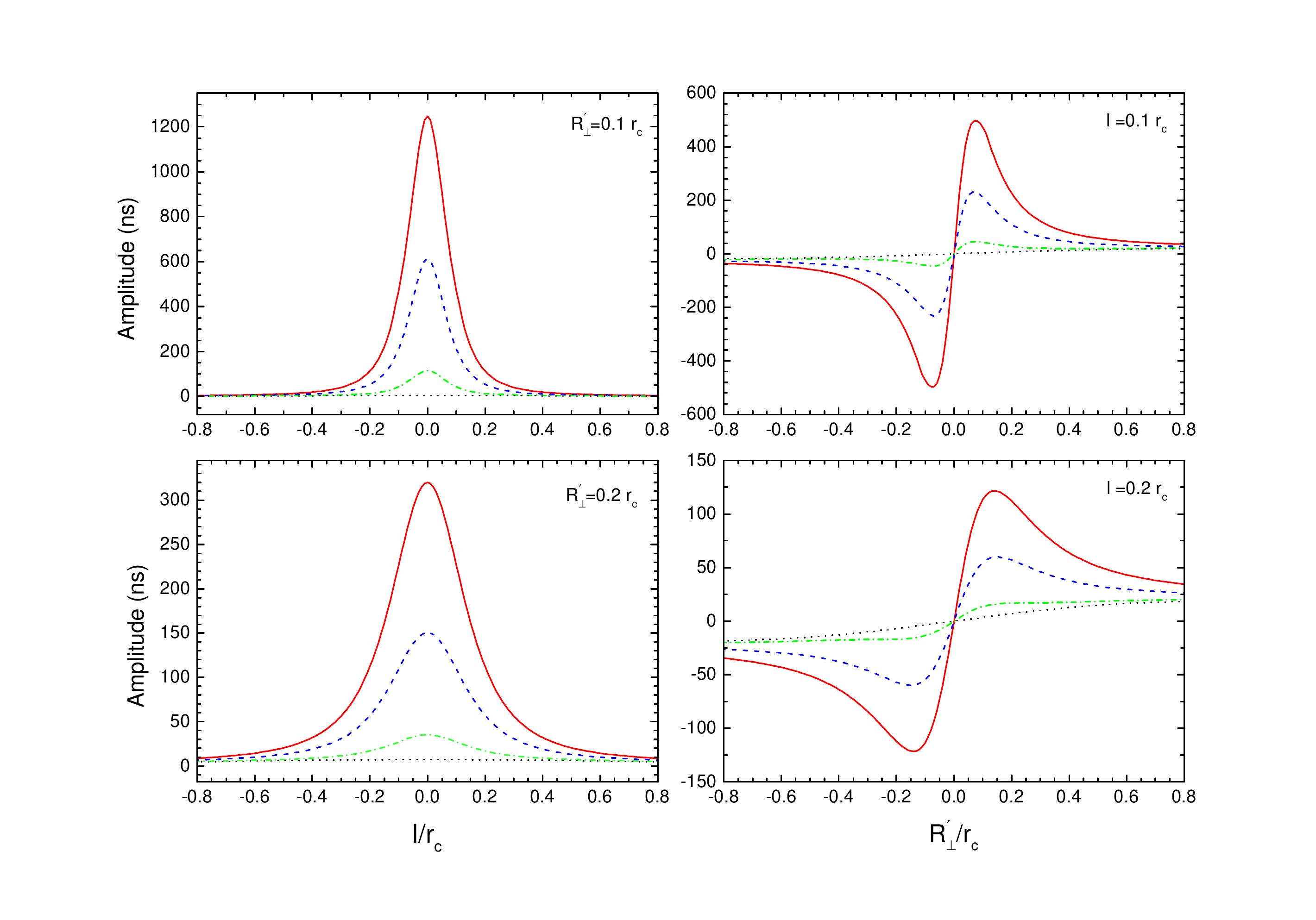}
\caption{The amplitudes of timing residuals due to the coupling effect of the pulsar transverse acceleration and the R$\rm{\ddot{o}}$mer delay, for pulsars in the inner region of a GC with $\rho_{c}=10^6~M_{\odot}{\rm pc}^{-3}$ and $r_{c}=0.2~{\rm pc}$, for a 20-yr observing campaign. Left panels: the residual amplitudes with respect to $l$, $R_{\perp}^{\prime}=0.1 r_{\rm c}$ (top panel) and $0.2 r_{\rm c}$ (bottom panel) are taken in the calculation. Right panel: the amplitudes with respect to $R_{\perp}^{\prime}$, $l=0.1 r_{\rm c}$ (top) and $0.2 r_{\rm c}$ (bottom) are taken. For all the panels, the dotted lines, dot-dashed lines, dashed lines, and solid lines represent for $m_{\rm BH}=0$, $1000$, $5000$, and $10000$, respectively.}\label{fig:1}
\end{figure*}

The amplitude of the timing residual due to the coupling effect of the transverse acceleration $\boldsymbol{a}_{\perp}$ and the R$\rm \ddot{o}$mer delay can be expressed as,
\begin{equation}\label{Coupling effect}
\Delta_{{\rm R}\odot}^\prime = \frac{\boldsymbol{a}_{\perp}\cdot\boldsymbol{r}_{\perp}}{2|\boldsymbol{R_{0}}| c}(t^{\rm psr}-t_{\rm pos})^2,
\end{equation}
where $c$ is the speed of light.

\begin{figure*}[!htbp]
\centering
\includegraphics[width=0.95\textwidth]{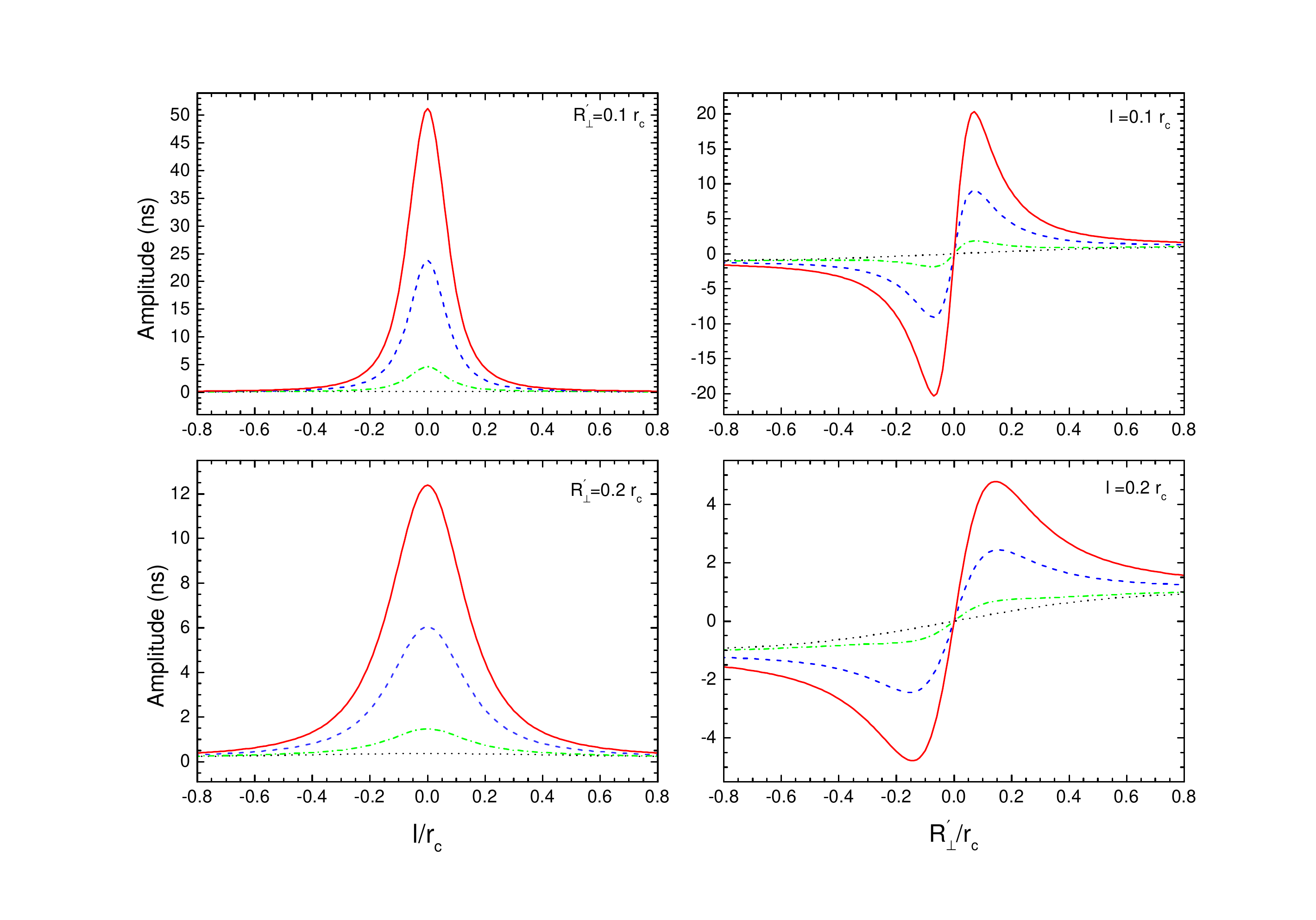}
\caption{The same as figure 1 but for pulsars in the inner region of a GC with $\rho_{c}=10^4~M_{\odot}{\rm pc}^{-3}$ and $r_{c}=1.0~{\rm pc}$.}\label{fig:2}
\end{figure*}

Due to the Doppler effect, a pulsar with rest frame period $P_0$ is observed to have a period $P$ \citep{1993ASPC...50..141P,2017ApJ...845..148P},
\begin{equation}\label{doppler}
P=[1+(\textbf{V}_{\rm p}-\textbf{V}_{\rm bary})\cdot \textbf{n}/c]P_0
\end{equation}
where $\textbf{V}_{\rm bary}$ is the velocity of the solar system barycenter, $\textbf{V}_{\rm p}$ is the pulsar velocity, and $\textbf{n}$ is the unit vector along the line-of-sight. The time derivative of equation (\ref{doppler}) gives \citep{2017ApJ...845..148P}
\begin{equation}\label{pdot}
\frac{\dot P}{P}=\frac{\dot P_0}{P_0}+\frac{a_{\rm l}}{c}+\frac{a_{\rm g}}{c}+\frac{a_{\rm s}}{c}+\frac{a_{\rm DM}}{c}
\end{equation}
in which the pulsar acceleration is decomposed into four terms, $a_{\rm l}$ is the line-of-sight acceleration due to the GC potential, $a_{\rm g}$ is the acceleration due to the Galactic potential, $a_{\rm s}$ is the apparent acceleration from the Shklovskii effect, and $a_{\rm DM}$ is the apparent acceleration due to errors in the changing dispersion measure. $a_{\rm g}$, $a_{\rm s}$, and $a_{\rm DM}$ are neglected in our model, since they are all much smaller than $a_{\rm l}$ \citep{2017ApJ...845..148P}.

We now derive the contribution for a pulsar's acceleration that arises due to the GC's mean field and the influence of the IMBH in the center of gravity (CoG)\footnote{It is assumed that the IMBH is fixed to the centre of the cluster, and this is most valid for larger black hole masses.}. We define a coordinate system for the GC: the plane passing through the CoG and perpendicular to our line-of-sight is defined as $O$, the impact parameter for a pulsar from the CoG as $R_{\perp}^{\prime}$, and the line-of-sight position going perpendicularly through $O$ as $l$, and thus the pulsar's spherical radius $r^{\prime}=\sqrt{R_{\perp}^{\prime 2}+l^2}$.

The IMBH for a given mass $M_{\rm BH}$ have a radius of influence \citep{2004ApJ...613.1133B},
\begin{equation}\label{influence radius}
r_{i}=\frac{3M_{\rm BH}}{8\pi \rho_{c}r_{c}^2},
\end{equation}
where $\rho_{c}$ is the core density of the GC, and $r_{c}$ is the core radius. Within $r_{i}$, the gravitational influence of the black hole is dominating, the density profile obeys $\rho_{\rm BH}\propto r^{-\alpha}$ \citep{2017ApJ...845..148P}. The power index $\alpha=1.55$ is found through N-body simulations of multiple component masses in the core \citep{2004ApJ...613.1143B}. At $r_{i}$ and beyond, the density profile follows the modified King density profile (Elson profile) as
\begin{equation}\label{King Model}
\rho(r^{\prime})\simeq\rho_{\rm c}[1+(r^{\prime}/r_{\rm c})^2]^{-\frac{\beta}{2}},
\end{equation}
in which the power-law slope $\beta$ is an undetermined parameter. The form of the density profile is widely used for star clusters in the Large Magellanic Cloud \citep{1992MNRAS.256..515E}, and the model back to King's model for $\beta=3$ \citep{1962AJ.....67..471K}. Integrating density profiles radially yields the interior mass at any given radius $r^{\prime}_{\ast}$. The acceleration felt at $r^{\prime}_{\ast}$ can be obtained by multiplying $G/r^{\prime 2}_{\ast}$, which reads,
\begin{equation}\label{Acceleration}
a(r^{\prime}_{\ast})=\frac{4\pi G}{r^{\prime 2}_{\ast}}[M_{\rm BH}/4\pi+\int_{0}^{r_{i}}r^{\prime 2}\rho_{\rm BH}dr^{\prime}+\int_{r_{i}}^{r^{\prime}_{\ast}}r^{\prime 2}\rho(r^{\prime})dr^{\prime}].
\end{equation}
The $\rho_{\rm BH}$ term contains only the density profile of GC stars under the influence of a central BH. For a radius beyond $r_{i}$, the BH has little impact on the density of GC stars. However, the acceleration felt by a pulsar should consider the gravity of the central BH (i.e. the first term in square bracket). One can get the transverse acceleration $a_{\perp}$ by projecting the acceleration $a(r^{\prime}_{\ast})$ along the transverse direction by a factor of $R_{\perp}^{\prime}/r^{\prime}_{\ast}$, or the line-of-sight acceleration $a_{\rm l}$ by a factor of $l/r^{\prime}_{\ast}$.

\section{Applications} \label{sec:applications}

\subsection{Timing Residuals due to The Coupling Effect}

The methods are applicable to all the galactic GCs. Among them, Ter5 has the largest number of identified MSPs, which is about a quarter of the total population ($39/230$) of pulsars in the galactic GCs\footnote{http://www.naic.edu/~pfreire/GCpsr.html}. The majority lies within the inner 20 arcminutes of the cluster \citep{2017ApJ...845..148P}. It is very suitable to use the basic parameters of Ter5 as an example for demonstrating the magnitudes of the coupling effect for pulsars distributing around the CoG. Using the accelerations and jerks of the ensemble of Ter5 MSPs, the core density $\rho_{c}=1.58^{+0.13}_{-0.13}\times 10^6~M_{\odot}{\rm pc}^{-3}$ and the core radius $r_{c}=0.16^{+0.01}_{-0.01}~{\rm pc}$ were obtained \citep{2017ApJ...845..148P}, and $r_{c}$ agrees with the values derived from high resolution \emph{Hubble Space Telescope} data \citep{2013ApJ...774..151M}. The most accurate distance estimate of the cluster is $d_0=5.9\pm 0.5~{\rm kpc}$ \citep{2007AJ....133.1287V}, and the value is inversely proportional to the residual amplitudes.\\

\begin{figure}[!htbp]
\centering
\includegraphics[width=0.60\textwidth]{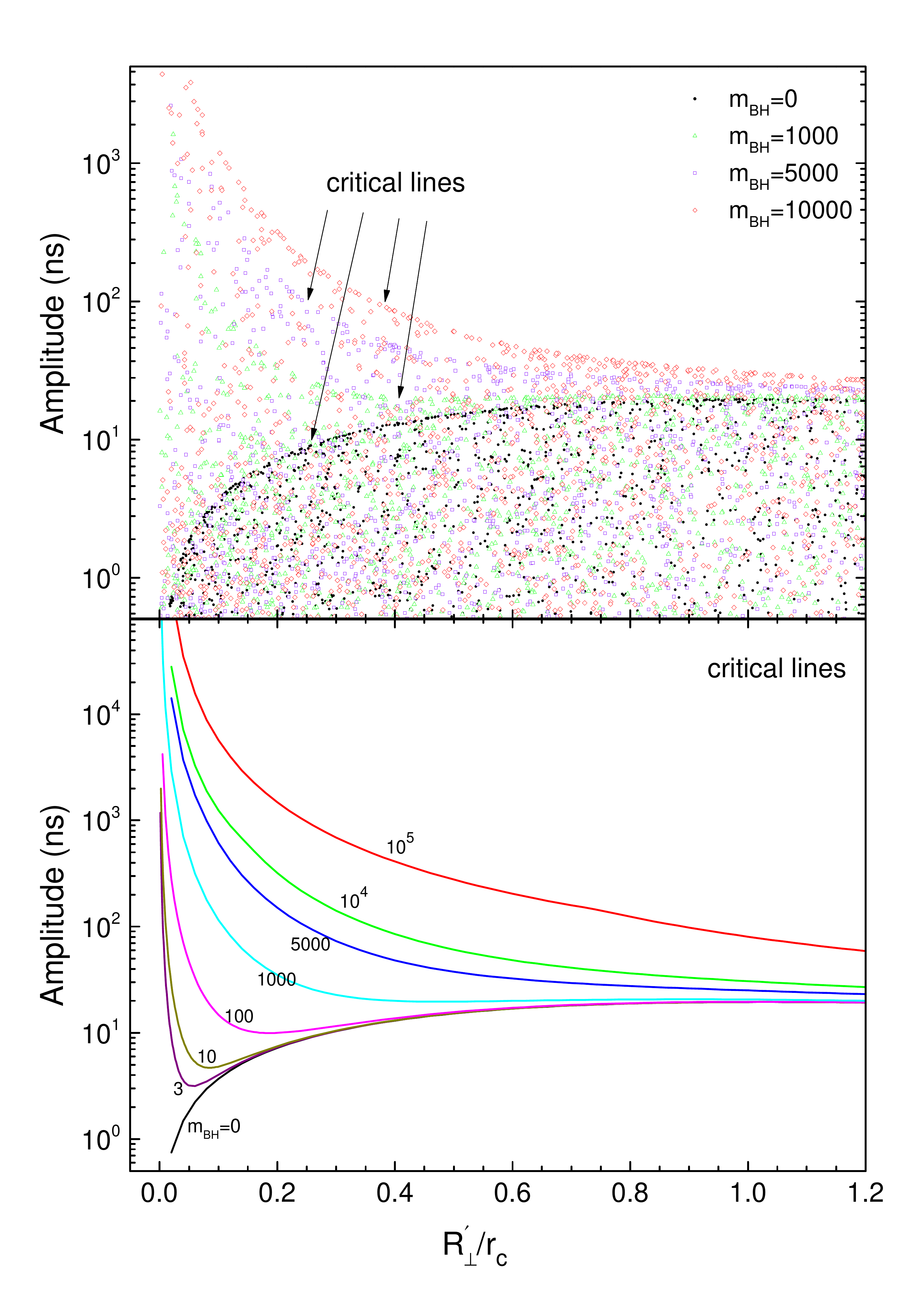}
\caption{\emph{Top}: the simulated distributions of residual amplitudes due to the coupling effect of the pulsar transverse acceleration and the R$\rm{\ddot{o}}$mer delay, for pulsars in the inner region of a GC with $\rho_{c}=10^6~M_{\odot}{\rm pc}^{-3}$ and $r_{c}=0.2~{\rm pc}$, for a 20-yr observing campaign. The black points, the green triangles, the blue squares, and the red diamonds represent for $m_{\rm BH}=0$, $1000$, $5000$, and $10000$, respectively. \emph{Bottom}: the critical lines of the distributions (from the maximum value fitting) for $m_{\rm BH}=0$, $3$, $10$, $100$, $1000$, $5000$, $10^4$, and $10^5$, respectively. }\label{fig:3}
\end{figure}

We consider a Ter5-like cluster with the core density $\rho_{c}=10^6~M_{\odot}{\rm pc}^{-3}$ and the core radius $r_{c}=0.2~{\rm pc}$, and a typical cluster with $\rho_{c}=10^4~M_{\odot}{\rm pc}^{-3}$ and $r_{c}=1.0~{\rm pc}$, and $d_0=10~{\rm kpc}$ is taken for both types in following calculations. We consider an IMBH with mass of $m_{\rm BH} M_{\odot}$ in the centre of the cluster. Combining equations (\ref{Coupling effect}), (\ref{influence radius})-(\ref{Acceleration}) and substituting the cluster parameters, one can obtain the amplitudes of the residuals for various $l$ or $R_{\perp}^{\prime}$. As examples, we show the residual amplitudes for pulsars in the Ter5-like cluster in figure \ref{fig:1}. The left panels are the amplitudes with respect to $l$ with $R_{\perp}^{\prime}=0.1 r_{\rm c}$ (top panel), and $0.2 r_{\rm c}$ (bottom panel). The right panels are the amplitudes with respect to $R_{\perp}^{\prime}$ with $l=0.1 r_{\rm c}$ (top) and $0.2 r_{\rm c}$ (bottom). For all the panels, the dotted lines, dot-dashed lines, dashed lines, and solid lines represent for $m_{\rm BH}=0$, $1000$, $5000$, and $10000$, respectively. The results imply that if there is an IMBH, the residuals due to the coupling effect may have some chances to be identified, particularly for those pulsars that distributed in the vicinity of the CoG and near the $O$ plane. We also calculated the case of no IMBH in the centre, the maximum amplitude is from the pulsar at the surface of the core, which is about $20$ nanoseconds (ns). This amplitude is relatively small, even though for those pulsars with highest timing precision \citep{2019MNRAS.490.4666P}. Figure \ref{fig:2} shows the residual amplitudes for pulsars in the inner region of the typical cluster with $\rho_{c}=10^4~M_{\odot}{\rm pc}^{-3}$ and $r_{c}=1.0~{\rm pc}$, the amplitudes are below $50$ ns for all the cases of $m_{\rm BH}\lesssim 10^4$ for pulsars beyond $0.1~r_{c}$, and thus hardly be observed currently. Cross calculations indicate that the amplitude differences between the Ter5-like cluster and the typical cluster are mainly due to the distance $r^{\prime}_{\ast}$ from the pulsar to the CoG, as well as the projection distance $R_{\perp}^{\prime}$, since at the inner most region of the clusters, the residuals are dominated by the effects of gravity from the IMBHs (if they are present), where the total mass of the interior stars $M_{\ast}$ is less than the mass of the IMBH in each cluster, e.g. $M_{\ast}< 100~M_{\odot}$ for $r^{\prime}_{\ast}<0.2 r_{\rm c}$ for both the two type of clusters.

For a complete profile of the residual magnitudes in the parameter space, we perform Monte Carlo (MC) simulations on the pulsar distribution in the core region of the cluster. In the simulations, the column number density profile of the pulsars is assumed as the following formula \citep{1995ApJ...439..191L}:
\begin{equation}\label{density profile}
n(x_{\perp})=n_0(1+x_{\perp}^2)^{q/2}.
\end{equation}
where $n_0$ is the central number density, $x_{\perp}\equiv R_{\perp}^{\prime}/r_{c}$, is the distance from the centre in the plane of the sky in units of the core radius, and $0<x\leq 10$ is taken in the following simulations. $q$ is the mass segregation parameter, we take the prior on this parameter to be a Gaussian centered on $-3$ with a dispersion of $0.5$ \citep{2019ApJ...884L...9A}. The mass interior to the radial position of each pulsar is calculated for the acceleration in simulations. Since the nearest neighbor stars present a negligible contribution to the line-of-sight acceleration \citep{2017ApJ...845..148P}, we propose that the transverse accelerations from the nearest neighbor stars are also smaller than that from the total interior mass. On the other hand, the residual patterns of the errors in the line-of-sight accelerations are different apparently from the patterns of the coupling effect. Thus their contributions on the residuals are also ignored, and only the residuals due to the coupling effect of the transverse acceleration from the total interior mass and the R$\rm{\ddot{o}}$mer delay are concerned here.

\begin{figure}[!htbp]
\centering
\includegraphics[width=0.5\textwidth]{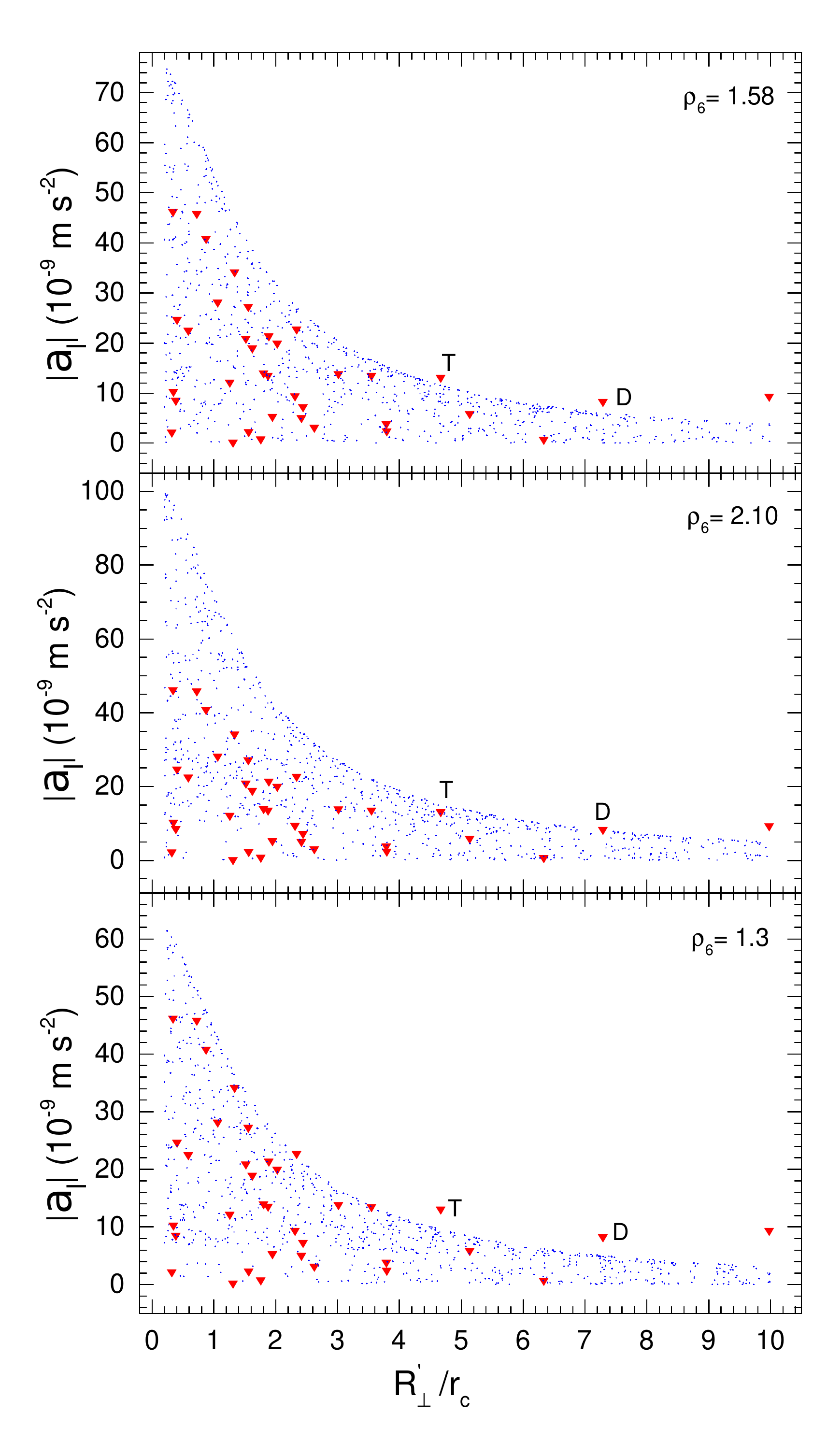}
\caption{The simulated $a_{\rm l}$ distributions for pulsars in Ter5. The red triangles represent for the measured $a_{\rm l}$ values, and the blue points are the simulated $a_{\rm l}$ values. The upper panel, middle panel and bottom panel show the simulations for $\rho_{6}=1.58$, $2.1$, and $1.3$, respectively. For all the simulated results, $r_{c}=0.16~{\rm pc}$ and $\beta=3$ are taken.}\label{fig:A}
\end{figure}

\begin{figure}[!htbp]
\centering
\includegraphics[width=0.5\textwidth]{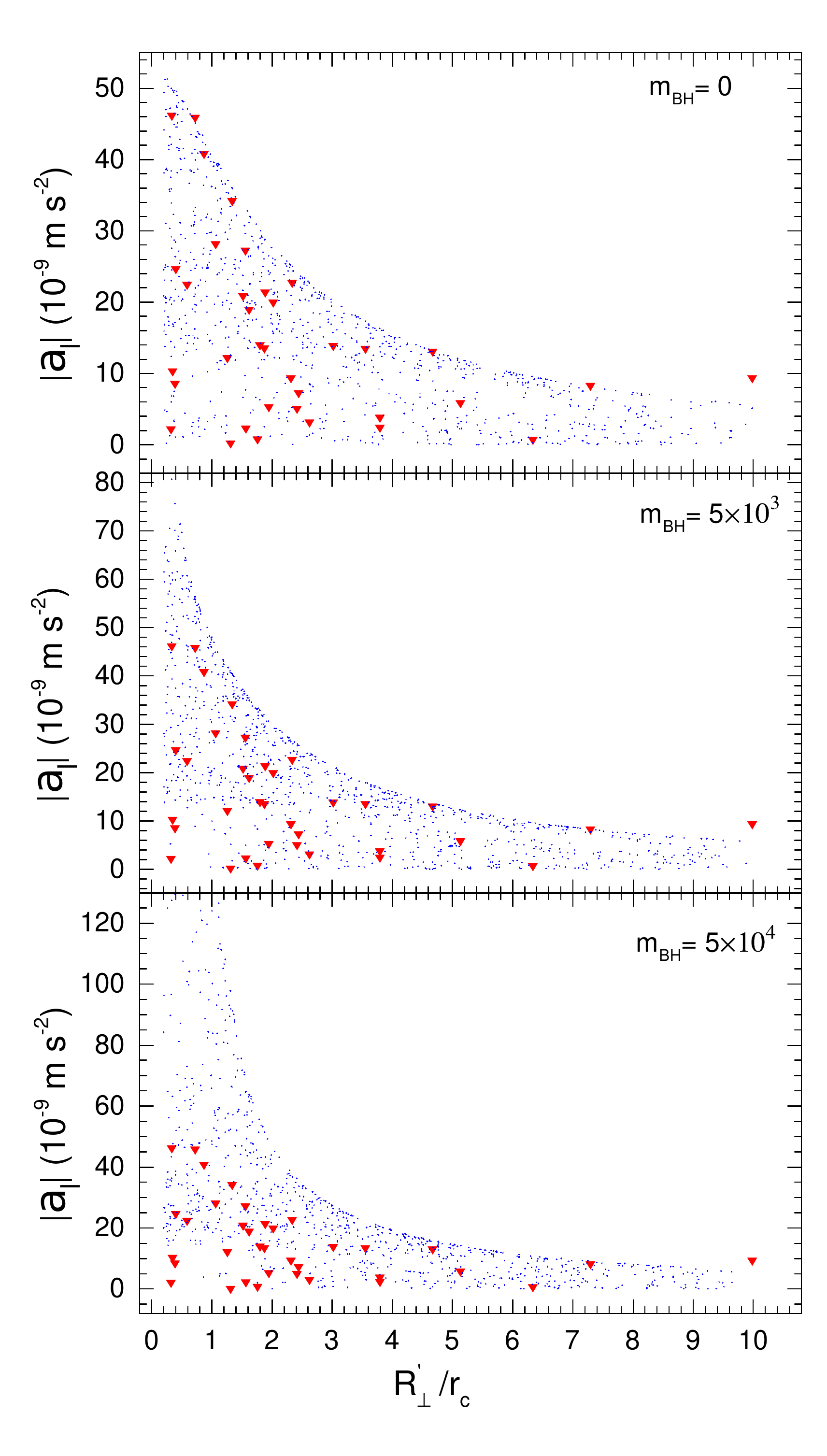}
\caption{The simulated $a_{\rm l}$ distributions for pulsars in Ter5. The red triangles represent for the measured $a_{\rm l}$ values, and the blue points are the simulated $a_{\rm l}$ values. The upper panel, middle panel and bottom panel show the simulations for $m_{\rm BH}=0$, $5\times 10^3$, and $5\times 10^4$, respectively. For all the simulated results, $\rho_{c}=0.95\times 10^6~M_{\odot}{\rm pc}^{-3}$, $r_{c}=0.16~{\rm pc}$, and $\beta=2.4$ are taken.}\label{fig:4}
\end{figure}

\begin{figure*}[!htbp]
\centering
\includegraphics[width=0.90\textwidth]{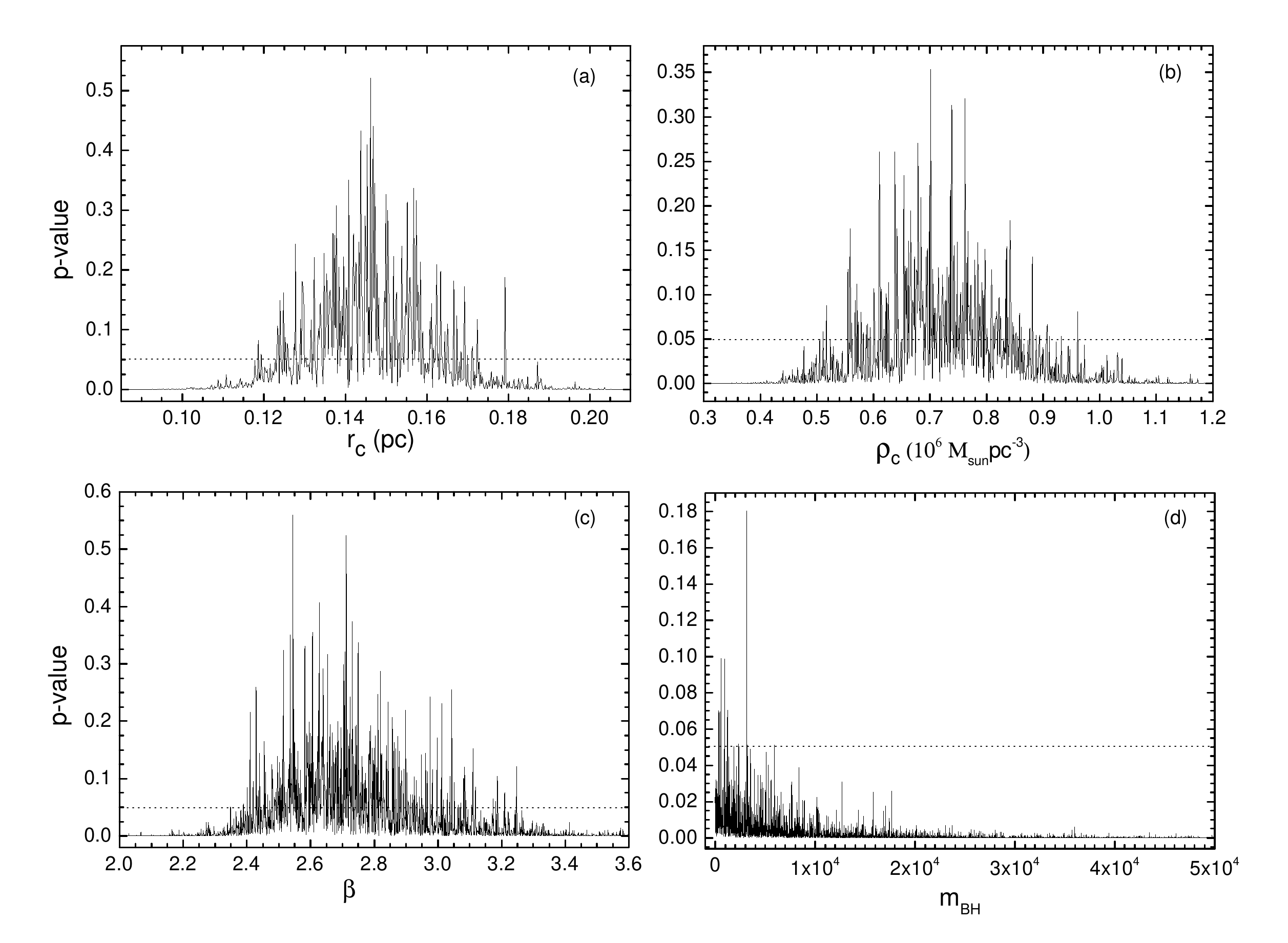}
\caption{The $p$-values of 2DKS for $R_{\perp}-a_{\rm l}$ distributions from the comparisons of measured data with simulated data. The ranges of $p$-values larger than 0.05 are identified by the transverse lines. Panel (a): for various values of $r_{\rm c}$. In the simulations, $\rho_{6}=0.95$, $\beta=2.4$, $m_{\rm BH}=0$ is taken. Panel (b): for various values of $\rho_{\rm c}$. In the simulations, $r_{\rm c}=0.16~{\rm pc}$, $\beta=2.4$, $m_{\rm BH}=0$ is taken.  Panel (c): for various values of $\beta$. In the simulations, $r_{\rm c}=0.16~{\rm pc}$, $\rho_{6}=0.95$, $m_{\rm BH}=0$ is taken.  Panel (d): for various values of $m_{\rm BH}$. In the simulations, $r_{\rm c}=0.16~{\rm pc}$, $\rho_{6}=0.95$, $\beta=2.4$ is taken.}\label{fig:5}
\end{figure*}

The comparison between the simulated results with the different mass of the central black holes are shown in the top panel of figure \ref{fig:3}, for the Ter5-like cluster, whose pulsar residuals could be much significant. The results for $m_{\rm BH}=0$, $1000$, $5000$ and $10^4$ are represented by the black points, the green triangles, the blue squares, and the red diamonds, respectively. The simulations agree with the results of figure \ref{fig:1}. Different from the case of $m_{\rm BH}=0$, the present of an IMBH in the centre strongly affects the residual magnitudes of pulsars around it. The residuals of some pulsars at the innermost region ($x_{\perp}\lesssim 0.4$) may be rather significant, and the root-mean-square of the residuals due to the effect may be higher than $1~\mu{\rm s}$. Up-to-date, the reported nearest pulsar from the centre of Ter5 on the plane of the sky is J$1748-2446$I, whose projected separation is about $0.3r_{c}$. The results imply that the high-magnitude residuals due to the coupling effect, may have chances to be identified for pulsars in the innermost region of GCs.

It is noteworthy that the residual amplitudes cannot be higher than the upper boundary line for $m_{\rm BH}=0$, unless an IMBH present. Thus the observations of pulsars with high-magnitude residuals will strongly support the existence of an IMBHs in the centre. The upper boundary line can actually play the role of a critical line for the black hole identification for each cluster. The critical lines can also be fitted with the maximum values of the simulated amplitudes, for different values of $m_{\rm BH}$. As shown in the bottom panel of figure \ref{fig:3}, a pulsar with residual amplitude higher than a critical line indicates the presence of a black hole with the mass higher than the corresponding value of the line. For symmetry reason we only show the region $R_{\perp}'/r_c \geqslant 0$ in figure 3, corresponding to the right half of the right panels of figure 1 and 2. In the right panels of figure 1 and 2, the amplitudes at $R_\perp'/r_c=0$ is exactly zero for a fixed $l$, and the peak values are larger and going to be close to $R_\perp'/r_c=0$ for a smaller $l$.  Therefore, the critical lines, which formed actually by all the maximum points of the amplitude lines of various values of $l$, approach to infinity when $l$ and $R_\perp'$ approaching to zero if an IMBH present.

\subsection{GC Parameter Fits Using $a_{\rm l}$ Data}

For the investigation of the structure parameters of Ter5, we perform Monte Carlo simulations on $a_{\rm l}$ distributions of pulsars in the inner region of the cluster with equations (\ref{pdot})-(\ref{density profile}). The timing parameters for Ter5 pulsars are taken from table 1 and the data of $R_{\perp}$ are taken from table 4 of \cite{2017ApJ...845..148P}. For convenience, we take $\rho_{6}=\rho_{c}/10^6~M_{\odot}{\rm pc}^{-3}$. The simulated distributions are shown in figure \ref{fig:A}. \textit{There are only a few pulsars with the maximum acceleration values that provide the strongest constraints for GC structure parameters. A precondition is assumed in the simulations that the area of the measured data should be well covered by the simulated data.} One can see that there are two pulsars, J1748-2446T and J1748-2446D, cannot be well covered with simulated data for the parameter $\rho_{6}=1.58$, as shown in the top panel. The coverage for the two pulsars needs $\rho_{6}\gtrsim 2.1$, as shown in the middle panel. However the coverage is excessive for the inner part for this case. The optimum coverage of the inner part requires $\rho_{6}\sim 1.3$, while the two pulsars cannot be covered else, as shown in the bottom panel.

In order to obtain the best-fit profile and turn measured $a_{\rm l}$ into a probe of the cluster potential,  we  substitute equation (\ref{King Model}) into equation (\ref{Acceleration}) in following calculations. We found that the simulated distribution with the parameters of $\rho_{c}=0.95\times 10^6~M_{\odot}{\rm pc}^{-3}$, $r_{c}=0.16~{\rm pc}$, $m_{\rm BH}=0$, and $\beta=2.4$ matches the measured data very well, as shown in the upper panel of figure \ref{fig:4}. Simulations for $m_{\rm BH}=5\times 10^3$ and $m_{\rm BH}=5\times 10^4$ are also shown in the middle and bottom panels, respectively. One can see that the IMBH can dramatically change the distributions of the most inner region.

It is very important to explore the parameter space of the simulations. We perform two-dimensional Kolmogorov-Smirnov (2DKS) test to reexamine the consistency of distributions of the simulated and measured $a_{l}$ for series values of the parameters under test. The 2DKS package\footnote{{http://www.downloadplex.com/Scripts/Matlab/Development-Tools/two-sample-two-diensional-kolmogorov-smirnov-test\b{~}432625.html}} is adopted for the test. Our strategy is then to search for the values of the parameters that can maximize the p-value of the 2DKS test against the hypothesis that the two distributions are consistent \citep{2019ApJ...880..123X}. We firstly let $r_{\rm c}$ vary from $0.08~{\rm pc}$ to $0.21~{\rm pc}$ with step size of $2\times 10^{-4}~{\rm pc}$. We draw 200 data points for each test, and the returned p-values are shown with solid lines in panel (a) of figure \ref{fig:5}. The p-value $0.05$, which is indicated by a dotted line, is considered as the threshold level with probability $95\%$. The test gives that $0.12 ~{\rm pc}\lesssim r_{c}\lesssim 0.19~{\rm pc}$. Similarly, the constraints on the parameters, $0.5 \lesssim \rho_{6}\lesssim 1.0 $ (testing from $0.3$ to $1.2$ with a step size of $10^{-3}$), and $2.4 \lesssim \beta\lesssim 3.2$ (testing from $2.0$ to $3.6$ with a step size of $10^{-3}$), are also obtained, and the results of p-values are shown in panel (b) and (c) of figure \ref{fig:5}, respectively. Considering the coverage condition, a tighter restriction on core density and $\beta$ can be made, $0.9 \lesssim \rho_{6}\lesssim 1.0 $ and $2.4\lesssim \beta \lesssim 2.6$.  We find good agreement with the result of \cite{2017ApJ...845..148P} for $r_{c}$, however, for $\rho_{c}$, there have an obvious difference. The main reason for the difference is probably the approximation of equation (27) of \cite{2017ApJ...845..148P}, since the obtained values of $a_{\rm {l,max}}$ with the equation (e.g. $a_{\rm {l,max}}\simeq 165$ $(10^{-9}~m~s^{-2})$ for $R_\perp'/r_c=1$, $r_{\rm c}=0.16~{\rm pc}$ and $\rho_{6}=1.67$) are apparently larger than the reported data shown in figure 4 and 5 of the manuscript. It is also noticed that the equation is actually different from equation (3.5) of \cite{1993ASPC...50..141P}. Unfortunately, a direct comparison with the latter is currently unavailable, as few values of velocity dispersion $\sigma$ have been reported in the region of $R_\perp'/r_c<10$ for the cluster \footnote{https://people.smp.uq.edu.au/HolgerBaumgardt/globular/}. Finally, we set $m_{\rm BH}$ vary from $0$ to $5\times 10^4$ with a step size of $5$, and $0< m_{\rm BH} \lesssim 6000$ is obtained. The results of p-values are shown in panel (d), which provide an upper limit for the mass of a possible IMBH at the core of the cluster. The mass segregation parameter $q$ is also tested. However, it is found that the method cannot place an effective restriction on $q$.

\section{Discussions and Conclusions} \label{sec:conclusions}

Pulsar timing residuals due to the coupling effect of the pulsar transverse acceleration and the R$\rm{\ddot{o}}$mer delay is usually negligible. For pulsars in the galactic field, the acceleration due to the galactic potential is of the order about $10^{-10}~{\rm m/s^2}$, which inducing timing residual $< 1 ~\rm{ns}$. Only for these pulsars in GCs, this effect is possibly needed. If no IMBH in the centre, the maximum amplitude gaining from the pulsars near surface of the core, is about tens of ns in a Ter5-like GC, thus hardly to be identified currently. However, an IMBH in the centre can increase apparently the residual magnitudes of pulsars in core region. The residuals of pulsars in the innermost region of GCs may be significant. The high-magnitude residuals, which above the critical lines of each cluster, are strong evidences for an IMBH in the centre. The timing effects of line-of-sight accelerations are also explored. The distributions of measured line-of-sight acceleration versus the projection radius are simulated with the MC method. The 2DKS tests are used to reexamine the consistency of distributions of the simulated and reported data for various values of parameters of the clusters. We found that the structure parameters of Ter5 can be well constrained by comparing the distributions of measured $a_{\rm l}$ with MC simulations. It is shown that Ter5 has an upper limit on the central black hole mass of $M_{\rm BH}\simeq 6000 M_{\odot}$. In the work, the same pulsar data from \cite{2017ApJ...845..148P} are used and a tighter constraint on IMBH mass ($\lesssim 6000 M_{\odot}$ compared to $\lesssim 30000 M_{\odot}$) is obtained. However, an intensive analysis for the improvement is still difficult to complete due to the complexity of the statistical processes.

Comparing with the Doppler effect (including the line-of-sight accelerations, the jerks and jounces), the residual from the coupling effect of the transverse acceleration and the R$\ddot{\rm o}$emer delay is relatively small, and has not been detected using pulsar timing, thus cannot provide tighter constraints on the masses of IMBHs currently. However, the residual whose magnitude is proportional to the observational time span, may probably be detected for pulsars in the innermost core region of GCs (see figure 3 of the manuscript) in the near future.  For a measured acceleration $a_l$, there are two possible line-of-sight positions ($l_1$ and $l_2$) that give the same acceleration, which causes a great uncertainty on the parameter determinations. Using the sign of a spin period derivative, one still cannot determine which of the two positions the pulsar is at, since they are on the same side of the pulsar, as shown in figure 1 of \cite{2017ApJ...845..148P}. However, the residual from the coupling effect can provide additional informations, i.e. constraints from the transverse component of the accelerations $a_{\perp}$, which may reduce greatly the uncertainty.

Many studies show that stellar-mass BHs could be common in the centers of GCs \citep{2013MNRAS.432.2779B, 2018MNRAS.479.4652A, 2020MNRAS.491..113H}. A group of stellar-mass BHs with total mass comparable to a single IMBH may have similar effects on the pulsar timing and accelerations. One may expect that the two values of $\beta$ that obtained from the surface brightness profiles or measured with pulsar accelerations should be the same for a single IMBH in the central. However, for the case of a population of stellar-mass BHs (or other dark remnants), the two values of $\beta$ may be apparently different in the central region of the GCs. Unfortunately, the current analysis cannot unambiguously discriminate between an IMBH and a group of stellar-mass BHs of comparable total mass.

We might expect some discoveries of the pulsars with the high-magnitude residuals in the future, as a good number of MSPs in GCs are reported and being timed regularly. We also expect to gain more details of the residuals and deeper understanding of GC dynamics, using future larger samples of MSPs with higher precision data, to be brought by China's Five-hundred-meter Aperture Spherical radio-Telescope (FAST) and the future Square Kilometre Array (SKA).\\

\begin{acknowledgements}
The anonymous referee is thanked for valuable comments and suggestions which helped to clarify several important points in the revised manuscript. This work is supported by National Natural Science Foundation of China under grant Nos. 11803009 and 11603009, and by the Natural Science Foundation of Fujian Province under grant Nos. 2018J05006, 2018J01416 and 2016J05013.

\end{acknowledgements}

\bibliography{ms}{}
\bibliographystyle{aasjournal}

\end{document}